\documentclass[a4paper,12pt]{revtex4}
\usepackage[latin1]{inputenc}
\usepackage{dsfont}
\usepackage{amsmath}
\usepackage{graphicx} %para poner graficos
\usepackage{amsfonts}
\usepackage{wasysym}
\usepackage{hyperref} %para poner automaticamente links a bibliografia y links a paginas web con \url{http://www.}

 \newcommand{\qed}{\hfill \ensuremath{\Box}}
\newtheorem{prop}{PROPOSITION}[section]
\newtheorem{defi}{DEFINITION}[section]
\newtheorem{corol}{COROLLARY}[section]
\newtheorem{conje}{CONJECTURE}[section]

\newtheorem{theo}{THEOREM}[section]

\begin{document}
\title{A new method to construct families \\of complex Hadamard matrices in even dimensions}
%\date{\today}
\author{D. Goyeneche}
\email{dgoyeneche@cefop.udec.cl}
\affiliation{Departamento de Fis\'{i}ca, Universidad de Concepci\'{o}n, Casilla 160-C, Concepci\'{o}n, Chile\\Center for Optics and Photonics, Universidad de Concepci\'{o}n, Casilla 4016, Concepci\'{o}n, Chile}
\begin{abstract}
We present a new method for constructing affine families of complex Hadamard matrices in every even dimension. This method has an intersection with Di\c{t}\u{a}'s construction and generalizes Sz\"oll\H{o}si's method. We extend some known families and present new ones existing in even dimensions. In particular, we find more than 13 millon inequivalent affine families in dimension 32. We also find analytical restrictions for any set of four mutually unbiased bases existing in dimension six and for any family of complex Hadamard matrices existing in every odd dimension.
\end{abstract}
\maketitle

\section{Introduction}
In recent years, the complex Hadamard matrices knowledge has exponentially increased. There are many applications to quantum information theory, e.g. they are useful to construct bases of unitary operators, bases of maximally entangled states and unitary depolarisers \cite{Werner}. Complex Hadamard matrices allow to solve the Mean King Problem \cite{Vaidman,Englert,Klappenecker}, to construct error correcting codes \cite{Heng}, to find quantum designs \cite{Zauner} and also to study spectral sets and Fuglede's conjecture \cite{Tao,Matolcsi,Kolountzakis,Kolountzakis2}. Furthermore, they are also useful for constructing some $*$-subalgebras in finite von Neumann algebras \cite{Popa,Harpe,Munemasa,Haagerup}, analyzing bi-unimodular sequences and finding cyclic $n$-roots \cite{Bjork,Bjork2} and equiangular lines \cite{Godsil}.

The existence of complex Hadamard matrices in every dimension is assured by the Fourier matrices. However, a complete characterization of inequivalent complex Hadamard matrices is known up to dimension five \cite{Haagerup}. The complexity of the problem suddenly increases in dimension six. This is not only due to the fact that six is not a prime power number, like what occurs in the mutually unbiased bases problem \cite{Wootters}. This issue remains open even in lower prime dimensions; e.g. it is open in dimension seven, where a one-parametric family and a few number of single complex Hadamard matrices are known. Also, we do not know if a continuous family exists in dimension eleven. A complete understanding of complex Hadamard matrices could help us to solve the Hadamard conjecture and the mutually unbiased bases problem in non-prime power dimensions.

In this work, we present a new method of constructing affine families of complex Hadamard matrices. This method allows us to find families stemming from a particular subset of complex Hadamard matrices existing in even dimensions. This subset includes the Fourier and real Hadamard matrices. This work is organized as follows: In Section II we briefly introduce complex Hadamard matrices. In Section III we present our method to generate affine families in even dimensions. We prove that inequivalent families of complex Hadamard matrices stem from inequivalent real Hadamard matrices when our method is used. We also find two interesting restrictions: \emph{(i)} for any set of four MU bases existing in dimension six and \emph{(ii)} for any family of complex Hadamard matrices existing in every odd dimension. In Section IV we compare our method with existing constructions; i.e. we generalize Sz\"oll\H{o}si's method and we demonstrate our method intersects Di\c{t}\u{a}'s construction. In Section V we exemplify our method by constructing families stemming from the Fourier matrices in every even dimension. In Section VI we construct families stemming from real Hadamard matrices. We also extend a known family in dimension eight and two families in dimension twelve. Finally, in Section VII we summarize and conclude.

\section{Complex Hadamard matrices}
In this section, we briefly resume basic properties of complex Hadamard matrices. More detailed explanations can be found in the book of K. Horadam \cite{Horadam} or in the self contained paper of W. Tadej and K. \.{Z}yczkowski \cite{Tadej2}. A square matrix $H$ of size $d$ is called a \emph{complex Hadamard matrix} if its entries are unimodular complex numbers and it has orthogonal columns. The Fourier matrix defined by its entries
\begin{equation}
    (F_d)_{j,k}=e^{\frac{2\pi i}{d}jk},
\end{equation}
where $i=\sqrt{-1}$ and $j,k=0,\dots,d-1$, is a privileged example because it represents the \emph{only} construction existing in every dimension $d$ \cite{Haagerup}. On the other hand, real Hadamard matrices, that is, complex Hadamard matrices having real entries, can only exist in dimensions of the form $d=4k$, where $k=1/2$ or $k$ is an integer positive number. The \emph{Hadamard conjecture} states that real Hadamard matrices exist in all such dimensions and it represents one of the most important open problems in Combinatorics. Currently, the smaller order where a real Hadamard matrix is still unknown is $d=4\times167=668$.

Two complex Hadamard matrices are \emph{equivalent} ($H_1\!\sim\! H_2$) if two diagonal matrices $D_1,D_2$ and two permutation matrices $P_1,P_2$ exist, such that
\begin{equation}
    H_2=D_1P_1H_1P_2D_2.
\end{equation}
In dimensions two, three and five every complex Hadamard matrix is equivalent to the Fourier matrix and in dimension four a uniparametric family stems from the Fourier matrix. This is the complete characterization of complex Hadamard matrices in $d\leq5$ \cite{Haagerup}. In dimensions higher than five a complete classification remains open. Recently, it has been proven that several four dimensional families exist in dimension six \cite{Dita5}. Strangely enough, its expressions are very complicated and none of them can be explicitly written in a single page.

A complex Hadamard matrix is \emph{dephased} if every entry of the first row and every entry of the first column are equal to the unity. Given a complex Hadamard matrix $H_1$ it is possible to obtain $H_2\sim H_1$ such that $H_2$ is written in dephased form and, conversely, if $H_1$ and $H_2$ have the same dephased form then $H_2\sim H_1$. However, the dephased form is not unique and we cannot use it to characterize inequivalent complex Hadamard matrices.

In the cases of $d=4$ and $d>5$ there exist continuous of inequivalent complex Hadamard matrices. This kind of sets is called a \emph{family}. A family is \emph{affine} \cite{Tadej2} if there exists a set $H(\mathcal{R})$ stemming from a dephased complex Hadamard matrix $H$, associated with a subspace $\mathcal{R}$ of the real space of $d\times d$ matrices with zeros in the first row and column such that
\begin{equation}
    H(\mathcal{R})=\{H\circ\exp(iR):R\in\mathcal{R}\}.
\end{equation}
Here, $R\in\mathbb{R}^{d^2}$ contains $m$ free parameters and generates an $m$-dimensional subspace with basis $R_1\dots,R_m$. We characterize the family with the notation $H(\vec{\xi})$, where $\vec{\xi}$ is an $m$-dimensional real vector. That is
\begin{equation}\label{affine}
    H(\vec{\xi})=H(R(\vec{\xi}))=H\circ\exp(iR(\vec{\xi})),
\end{equation}
where $R(\vec{\xi})=\sum_{i=1}^m\xi_iR_i$. The symbol $\circ$ denotes the Hadamard product
\begin{equation}
    (H_1\circ H_2)_{ij}=(H_1)_{ij}(H_2)_{ij},
\end{equation}
while $\mathrm{Exp}$ denotes the entrywise exponential function
\begin{equation}
    (\mathrm{Exp}(H))_{ij}=\exp(H_{ij}).
\end{equation}
We say an affine family is \emph{maximal} if it is not contained in any larger affine family $H(R')$ stemming from $H$, where $\{R\}\subset \{R'\}$. If a family cannot be written in the form of Eq.(\ref{affine}) we say it is \emph{non-affine}. By the other hand, if a complex Hadamard matrix does not belong to a family it is \emph{isolated}. For example, the spectral matrix $S_6$ and every Fourier matrix defined in prime dimensions are isolated \cite{Tadej}. The transpose of a family is still a family. We say that two families $H_1(\vec{\xi})$ and $H_2(\vec{\nu})$ are \emph{cognate} if the families $H_2(\vec{\nu})$ and $H_1^t(\vec{\xi})$ are equivalent. Here, $t$ denotes matrix transposition. If a family and its transpose determine the same family we say it is \emph{self-cognate}. For example, the one-parametric family $F_4^{(1)}$ is self-cognate.

The problem of determining the maximal family stemming from a complex Hadamard matrix is open in dimensions higher than five. Even the problem to find the dimension of the maximal family is still open. The best approach in order to find this number is the \emph{defect} of a complex Hadamard matrix \cite{Tadej}, that is, the dimension of the solution space of the linear system
\begin{equation}
    \left\{ \begin{array}{rl}
R_{0,j}=0,&\hspace{0.5cm}j\in\{1,\dots,d-1\},\\
R_{i,0}=0,&\hspace{0.5cm}i\in\{0,\dots,d-1\},\\
\sum_{k=0}^{d-1}H_{i,k}H_{j,k}^*(R_{i,k}-R_{j,k})=0,&\hspace{0.5cm}0\leq i<j\leq d-1,
 \end{array} \right.
\end{equation}
where $R$ is a variable matrix. The defect $\mathbf{d}(H)$ is an upper bound of the dimension of the maximal family stemming from $H$. For example, $\mathbf{d}(H)=0$ implies that $H$ is an isolated matrix but the reciprocal implication is not valid. The defect has been analytically obtained for the Fourier matrices $F_d$ in every dimension \cite{Tadej}. As a particularly interesting case, it has been proven that $\mathbf{d}(F_d)=0$ when $d$ is prime. Consequently, the Fourier matrices in prime dimensions are isolated. Sometimes, this upper bound is not attained: $\mathbf{d}(F_4)=1$ but $\mathbf{d}(F_2\otimes F_2)=3$, and the maximal affine family existing in dimension four is one-dimensional only \cite{Haagerup}.

\section{Construction of affine families}\label{construction}
A family of complex Hadamard matrices defines a continuous set of orthogonal bases in $\mathbb{C}^d$ when the parameters of the family are smoothly changed. These bases are given by the columns of such matrices which rotate in a very special way; i.e. orthogonality is preserved and also every entry of every column is restricted to be a unimodular complex number. In this section, we deal with a very particular kind of rotations. The main idea of this work comes from the following question:
\begin{center}
\emph{Can we define a family of complex Hadamard matrices by introducing\\ a parameter in two columns of a single complex Hadamard matrix?}
\end{center}
As we will show, this question has only a positive answer for even dimensions. Let $\{\phi_k\}$ be an orthogonal base such that every vector $\phi_k$ defines a column of a $d\times d$ complex Hadamard matrix $H$. Our objective consists in finding two continuous vectors $\phi_a(\xi)$ and $\phi_b(\xi)$ such that:
\begin{enumerate}
   \item[(C.1)] \emph{They are a linear combination of} $\phi_0$ \emph{and} $\phi_1$.
   \item[(C.2)] \emph{They have unimodular complex entries.}
   \item[(C.3)] \emph{They are orthogonal.}
   \item[(C.4)] \emph{The initial conditions} $\phi_a(0)=\phi_0$ \emph{and} $\phi_b(0)=\phi_1$ \emph{hold}.
\end{enumerate}
If these conditions are satisfied then the vectors $\{\phi_a(\xi),\phi_b(\xi),\phi_2,\dots,\phi_{d-1}\}$ define a family of complex Hadamard matrices. Let $\{\varphi_k\}$ be the canonical base and let $\{\phi_0,\phi_1\}$ be the first two columns of $H$. Without loosing the generality we can assume that
\begin{equation}
    \phi_0=\sum_{k=0}^{d-1}\varphi_k,
\end{equation}
and
\begin{equation}
    \phi_1=\sum_{k=0}^{d-1}e^{i\alpha_k}\varphi_k,
\end{equation}
for a given set of unimodular complex numbers $\{e^{i\alpha_k}\}$ restricted to the following condition
\begin{equation}\label{restr2}
    \sum_{k=0}^{d-1}e^{i\alpha_k}=0.
\end{equation}
Therefore, proposing a linear combination of $\phi_0$ and $\phi_1$ (C.1)
\begin{eqnarray}\label{phi_a}
    \phi_a(\xi)&=&x(\xi)\phi_0+y(\xi)\phi_1,\nonumber \\
             &=&\sum_{k=0}^{d-1}(x(\xi)+y(\xi)e^{i\alpha_k})\varphi_k,
\end{eqnarray}
and imposing unimodular entries in the last equation (C.2) we obtain
\begin{equation}\label{restr1}
    |x(\xi)+y(\xi)e^{i\alpha_k}|=1,
\end{equation}
for every $k=0,\dots,d-1$. The coupled system of equations given by Eqs.(\ref{restr2}) and (\ref{restr1}) has a solution if and only if $d$ is an even number and $\phi_1$ is a real vector. In fact, expanding Eq.(\ref{restr1}) we obtain
\begin{equation}\label{restr1_2}
    |x(\xi)|^2+|y(\xi)|^2+2\mathrm{Re}(x^*(\xi)y(\xi)e^{i\alpha_k})=1,
\end{equation}
for every $k=0,\dots,d-1$. Given that $\phi_a(\xi)$ defined in Eq.(\ref{phi_a}) is normalized, that is
\begin{equation}\label{rest1_3}
    |x(\xi)|^2+|y(\xi)|^2=1,
\end{equation}
we have
\begin{equation}\label{rest1_4}
    \mathrm{Re}(x^*(\xi)y(\xi)e^{i\alpha_k})=0,
\end{equation}
for every $k=0,\dots,d-1$ and $\xi\in[0,2\pi)$. One way to write the most
general solution of Eq.(\ref{rest1_3}) is
\begin{equation}
    x(\xi)=\cos(\xi)\hspace{0.3cm}\mbox{and}\hspace{0.3cm}y(\xi)=\sin(\xi)e^{i\beta(\xi)},
\end{equation}
and imposing Eq.(\ref{rest1_4}) we obtain that
\begin{equation}\label{phase1}
    e^{i\beta(\xi)}e^{i\alpha_k}=(-1)^ki,
\end{equation}
for every $\xi\in[0,2\pi)$. Therefore,
\begin{equation}\label{sol1}
    \phi_a(\xi)=\sum_{k=0}^{d-1}e^{i(-1)^k\xi}\varphi_k,
\end{equation}
up to equivalence. The only pure state that is orthogonal to $\phi_a(\xi)$ (C.3) and it is also a linear combination of $\phi_0$ and $\phi_1$ (C.1) is given by
\begin{equation}
    \phi_b(\xi)=y(\xi)^*\phi_0-x(\xi)\phi_1.
\end{equation}
From Eq.(\ref{phase1}) without loosing the generality we can choose
\begin{equation}
    e^{i\beta(\xi)}=i\hspace{0.3cm}\mbox{and}\hspace{0.3cm}e^{i\alpha_k}=(-1)^k,
\end{equation}
and up to a global sign we obtain
\begin{equation}\label{sol2}
    \phi_b(\xi)=\sum_{k=0}^{d-1}(-1)^ke^{i(-1)^k\xi}\varphi_k.
\end{equation}
From the last equation we show that the entries of $\phi_b(\xi)$ are unimodular complex numbers (C.2) and also that the initial conditions imposed in (C.4) hold. Note that the only difference between $\{\phi_0,\phi_1\}$ and $\{\phi_a(\xi),\phi_b(\xi)\}$ is the exponential term $\{e^{i(-1)^k\xi}\}$ appearing in Eqs.(\ref{sol1}) and (\ref{sol2}). We highlight this is the most general solution up to equivalence. Another interesting consequence arises from Eqs.(\ref{sol1}) and (\ref{sol2}): \emph{our construction only works in even dimensions}. Otherwise, the vectors $\phi_a(\xi)$ and $\phi_b(\xi)$ are not orthogonal. Before formalizing the above results in a theorem let us define a useful concept.
\begin{defi}
Let $C_A$ and $C_B$ be two columns of a complex Hadamard matrix. We say they are an equivalent to real (ER) pair if
\begin{equation}
    (C_A^*)_j(C_B)_j=\pm1,
\end{equation}
for every $j=0,\dots,d-1$. Here, $(C_A)_j$ and $(C_B)_j$ are the $j$th entries of $C_A$ and $C_B$, respectively. The asterisk denotes complex conjugation.
\end{defi}
For example, the Fourier matrix defined in every even dimension $d$ has $d/2$ ER pairs of columns. Indeed, the $k$th column of $F_d$ is given by
\begin{equation}
    (F_d)_k=\sum_{l=0}^{d-1}\omega^{lk}\varphi_l,
\end{equation}
where $\omega=e^{2\pi i/d}$. The $d/2$ ER pairs of $F_d$ are determined by
\begin{equation}\label{ERpairs}
    \left\{(F_d)_k,(F_d)_{k+\frac{d}{2}}\right\},
\end{equation}
where $k=0,\dots,\frac{d}{2}-1$. Given that $F_d$ is symmetric it also has $d/2$ ER pairs of rows. The maximal number of ER pairs of columns and rows are denoted by $\eta_c$ and $\eta_r$, respectively. These numbers coincide for the Fourier matrices but they differ in general. For example, the spectral matrix $S_8$ \cite{Matolcsi2} has $\eta_c=4$ and $\eta_r=0$. We formalize the results found in Eqs.(\ref{sol1}) and (\ref{sol2}) in the following theorem:
\begin{theo}\label{prop1}
Let $H$ be a complex Hadamard matrix defined in an even dimension $d>2$. If $H$ has $m<d/2$ ER pairs then $H$ belongs to a $m$-dimensional family.
\end{theo}
If we consider $d/2$ ER pairs of columns one of the $d/2$ parameters generated by our method could be linearly dependent on the others. For example, this occurs for the family stemming from the Fourier matrix in every even dimension. Let us analyze in a separated subsection all consequences of Theorem \ref{prop1}.

\subsection{Consequences of Theorem \ref{prop1}}
Theorem \ref{prop1} is the main result of this paper and it has several consequences. We have collected all of them in this subsection in order to have a clear structure of our results. The following corollaries emerge from the above theorem
\begin{corol}\label{nonisolated}
A complex Hadamard matrix having an ER pair of columns or rows is not isolated.
\end{corol}
\begin{corol}\label{oddcorol}
A parameter of a family of complex Hadamard matrices defined in every odd dimension cannot appear in only two columns or rows.
\end{corol}
The proof of these corollaries is trivial from Theorem \ref{prop1}. Our intention here is to emphasize that \emph{(i)} two columns of $H$ can contain enough information to affirm that a complex Hadamard matrix \emph{is not} isolated and \emph{(ii)} our construction forbids its extension to every odd dimension. Let us define a particularly interesting case of ER pairs.
\begin{defi}
Let $\{C_1,C_2\}$ and $\{C_3,C_4\}$ be two ER pairs of columns. We say that they are aligned if
\begin{equation}
    (C_1^*)_k(C_2)_k=(C_3^*)_k(C_4)_k,
\end{equation}
for every $k=0,\dots,d-1$. The maximal number of aligned pairs of columns and rows is called $\eta_{\bar{c}}$ and $\eta_{\bar{r}}$, respectively.
\end{defi}
Let us prove that the existence of ER pairs is invariant under equivalence.
\begin{prop}\label{propequi}
Let $H$ and $\tilde{H}$ be two equivalent and dephased complex Hadamard matrices. Let $\{C_1,C_2\}$ be an ER pair of columns of $H$. Then, the corresponding pair $\{\tilde{C_1},\tilde{C_2}\}$ is an ER pair of columns of $\tilde{H}$.
\end{prop}
Proof:
Let $\{C_1,C_2\}$ an ER pair of columns of $H$. Hence, up to equivalence this pair is given by
\begin{equation}
    C_1=\sum_{k=0}^{d-1}a_k\varphi_k,
\end{equation}
and
\begin{equation}\label{C2}
    C_2=\sum_{k=0}^{d-1}(-1)^ka_k\varphi_k.
\end{equation}
Therefore, we have
\begin{equation}\label{aligned1}
    (C_1)_k(C_2)_k=(-1)^k.
\end{equation}
Let $\tilde{H}$ be a complex Hadamard matrix equivalent to $H$. Therefore, there exists unimodular complex numbers $c$ and $b_k$, and an injective function $f:\mathbb{Z}_d\rightarrow\mathbb{Z}_d$ such that
\begin{equation}\label{equi1}
    \tilde{C_1}=\sum_{k=0}^{d-1}c\,b_{k}\,a_{f(k)}\,\varphi_{k},
\end{equation}
and
\begin{equation}\label{equi2}
    \tilde{C_2}=\sum_{k=0}^{d-1}(-1)^{f(0)}\,c\,b_{k} \,(-1)^{f(k)}a_{f(k)}\,\varphi_{k}.
\end{equation}
The function $f(k)$ and the numbers $b_k$ are related to a permutation operator $P$ and a diagonal unitary operator $D$ applied to $H$, respectively. In order to dephase $\tilde{C_1}$ and $\tilde{C_2}$ we consider
\begin{equation}
c=(b_{0}\,a_{f(0)})^*.
\end{equation}
From Eqs.(\ref{equi1}) and (\ref{equi2}) we obtain
\begin{equation}\label{aligned2}
    (\tilde{C_1})^*_k(\tilde{C_2})_k=(-1)^{f(0)}\,(-1)^{f(k)}.
\end{equation}
Therefore, $\tilde{C_1}$ and $\tilde{C_2}$ are an ER pair of columns.\qed

We remark that we could consider $(-1)^{g(k)}$ instead of $(-1)^k$ in Eq.(\ref{C2}), where ${g:\mathbb{Z}_d\rightarrow\mathbb{Z}_d}$ is an injective function. Nevertheless, our consideration is general up to equivalence. Let us present the main consequence of this proposition.
\begin{corol}
Let $H$ be a dephased complex Hadamard matrix. Then, the numbers $\eta_c,\eta_r,\eta_{\bar{c}}$ and $\eta_{\bar{r}}$ are invariant under equivalence.
\end{corol}
Proof: The numbers $\eta_c$ and $\eta_r$ are trivially invariant from Proposition \ref{propequi}. On the other hand, $\eta_{\bar{c}}$ is invariant due to the function $f$ appearing in Eq.(\ref{aligned2}) which is the same for every ER pair of columns; analogously for $\eta_{\bar{r}}$.\qed

In the case of the Fourier matrices we have
\begin{eqnarray}
\eta_c= \left\{
\begin{array}{c l}
\frac{d}{2} &\mbox{if $d$ is even,}\\
0           &\mbox{if $d$ is odd,}
\end{array}
\right.
\end{eqnarray}
and the same values for $\eta_r,\eta_{\bar{c}}$ and $\eta_{\bar{r}}$. However, for real Hadamard matrices $\eta_c=\eta_r=d/2$ but, it is not possible to guess the functions $\eta_{\bar{c}}$ and $\eta_{\bar{r}}$ \emph{a priori}. This is due to the existence of inequivalent real Hadamard matrices in dimensions $d\geq16$. Let us present some properties of ER pairs in dimension six.
\begin{prop}
A $6\times6$ complex Hadamard matrix $H$ belongs to the family $F_6^{(2)}$ if and only if $\eta_c\neq0$ or $\eta_r\neq0$.
\end{prop}
Proof: If $H$ belongs to $F_6^{(2)}$ it is trivial to prove that it has 3 ER pairs of columns and rows. This is because the number of ER pairs is the same for any parameter of the family, as we can see from Eqs.(\ref{sol1}) and (\ref{sol2}). Therefore, $\eta_c=\eta_r=3$ for any member of the family $F_6^{(2)}$. Reciprocally, let us suppose that $H$ is written in a dephased form and $\eta_c\neq0$. Therefore, there exists $\tilde{H}\sim H$ such that it has a real ER pair of columns. Considering that every entry of $\tilde{H}$ is unimodular and every pair of columns is orthogonal we only obtain two free parameters. These parameters generate $F_6^{(2)}$. Analogously for $\eta_r\neq0$.\qed

Given a pair of MU bases $\{\mathbb{I},F_6^{(2)}(a,b)\}$ it is not possible to find more than triplets of MU bases for any parameters $(a,b)$ \cite{Jaming}. From this fact and the last proposition an interesting consequence emerges.
\begin{corol}\label{CorolMUB}
Suppose that four MU bases can be constructed in dimension six, namely $\{\mathbb{I},H_1,H_2,H_3\}$. Then, $\eta_c=\eta_r=0$ for $H_1,H_2$ and $H_3$.
\end{corol}
This corollary means that if such matrices $H_1,H_2,H_3$ exist, each of them is inequivalent to a complex Hadamard matrix having two real columns. This new result is one of the very few analytically known restrictions for the existence of four MU bases in dimension six.

Let us consider ER pairs in dimension four. Here, the most general ER pair of columns can be written in the form
\begin{eqnarray}
    C_1&=&(1,e^{ia},e^{ib},e^{ic}),\\
    C_2&=&(1,-e^{ia},e^{ib},-e^{ic}),
\end{eqnarray}
for any numbers $a,b,c\in[0,2\pi)$. The $-1$'s can appear in other entries but the above case is general up to equivalence. The inner product between these columns is
given by
\begin{equation}\label{pm}
    \langle C_1,C_2\rangle=1-1+1-1=0.
\end{equation}
As we can see, the contribution of every pair of entries to the inner product is $\pm1$; this is an exclusive property of ER pairs. Using Eqs.(\ref{sol1}) and (\ref{sol2}) we construct the continuous pair of ER vectors
\begin{eqnarray}
    C_1(\xi)&=&(e^{i\xi},e^{i(a-\xi)},e^{i(b+\xi)},e^{i(c-\xi)}),\\
    C_2(\xi)&=&(e^{i\xi},-e^{i(a-\xi)},e^{i(b+\xi)},-e^{i(c+\xi)}).
\end{eqnarray}
Dephasing and considering $\xi\rightarrow-\xi/2$ we find that
\begin{eqnarray}
    C_1(\xi)&=&(1,e^{i(a+\xi)},e^{ib},e^{i(c+\xi)}),\\
    C_2(\xi)&=&(1,-e^{i(a+\xi)},e^{ib},-e^{i(c+\xi)})\label{C1C2}.
\end{eqnarray}
Note that $\xi$ only appears in entries such that $(C_1)^*_j(C_2)_j=-1$. This is a \emph{mnemonic technique} to construct families from our method. Let us present a proposition regarding families stemming from real Hadamard matrices.
\begin{prop}\label{propHad}
Let $H_1$ and $H_2$ be two inequivalent real Hadamard matrices. Then, the families stemming from them by using our method are inequivalent.
\end{prop}
Proof: Let $H_1(\vec{\xi}\,)$ be a family stemming from the real Hadamard matrix $H_1$ and constructed from our method. Then, the only way to obtain a real Hadamard matrix into $H_1(\vec{\xi}\,)$ is by considering every entry of $\vec{\xi}$ in the set $\{0,\pi\}$. By the other hand, from Eqs.(\ref{sol1}) and (\ref{sol2}) it is easy to show that $H_1(\vec{\xi_1}\,)\sim H_1(\vec{\xi_2}\,)$ when $\vec{\xi}_1$ and $\vec{\xi}_2$ have every entry in the set $\{0,\pi\}$. Moreover, every vector contained in an ER pair is changed, at most, in a global sign under these considerations. Therefore, a real Hadamard matrix $H_2\not\sim H_1$ cannot be contained in the family $H_1(\xi\,)$.\qed

Despite of Proposition \ref{propHad}, $H_1\sim H_2$ does not imply that the families stemming from these matrices are equivalent, as we can see in Di\c{t}\u{a}'s family $D^{(7)}_{12\Sigma}$ \cite{Dita4}.
\begin{corol}\label{Corol13millon}
In dimension 32 we can generate 13,710,027 inequivalent affine families of complex Hadamard matrices.
\end{corol}
This corollary is a consequence of the existence of exactly 13,710,027 inequivalent real Hadamard matrices in dimension 32 \cite{Kharaghania}. Additionally, in dimensions $16, 20, 24$ and $28$ we can construct 5, 3, 60 and 487 inequivalent affine families, respectively. Let us present some general cases where our construction cannot be applied.
\begin{prop}
Let $H$ be a dephased $d\times d$ complex Hadamard matrix such that one of the following conditions hold
\begin{enumerate}
  \item The entries of $H$ are real in the main diagonal and non-real in other cases.
  \item The entries of $H$ are power of roots of the unity $\omega^k$ with $k=0,\dots,d/2-1$, where $\omega=e^{\frac{2\pi i}{d}}$.
  \item The entries of $H$ are power of roots of the unity $\omega=e^{\frac{2\pi i}{N}}$, where $N$ is an odd number.
\end{enumerate}
Then, our method defined in Theorem \ref{prop1} cannot be applied.
\end{prop}
The proof is trivial because in these cases we do not have ER pairs.\qed

For example, we cannot construct a family stemming from the matrices $C_6,D_6,D_{10}$ and $D_{14}$ (see the BTZ catalog \cite{Bruzda} for explicit expressions of these matrices). Also, this proposition tells us that we cannot construct a family stemming from the isolated matrix $S_6$ and from any isolated matrix recently found by McNulty and Weigert \cite{McNulty}.

\section{Sz\"oll\H{o}si and Di\c{t}\u{a} methods}\label{Szol_Dita}
In dimension four or higher than five several families of complex Hadamard matrices have been found. Some of them are specific constructions and cannot be extended to other dimensions. A method found by Sz\"oll\H{o}si allows us to find a family of complex Hadamard matrices stemming from real Hadamard matrices. This method is stated in Lemma 3.4 \cite{Szollosi}:

\textbf{Lemma 3.4} \emph{(Sz\"oll\H{o}si)}\hspace{0.2cm}\emph{Let $H$ be an arbitrary dephased complex Hadamard matrix of order $d\geq 4$. Suppose that $H$ has a pair of columns, say $u$ and $v$, with the following property: $u_i = v_i$ or $u_i + v_i = 0$ holds for every $i =0,\dots,d-1$. Then, $H$ admits an affine orbit.}

This lemma is complemented by Theorem 3.5, which represents the main result of Sz\"oll\H{o}si's paper:

\textbf{Theorem 3.5} \emph{(Sz\"oll\H{o}si)}\hspace{0.2cm} \emph{Let $H$ be a real Hadamard matrix of order $d\geq12$. Then, $H$ admits an $d/2+1$-parameter affine orbit.}

We have noticed that Lemma 3.4 coincides with our Corollary \ref{nonisolated}. It is important to realize that the main result found by Sz\"oll\H{o}si is Theorem 3.5, which considers families stemming from real Hadamard matrices. This theorem provides the first proof that any real Hadamard matrix defined in $d\geq12$ is not isolated. However, Lemma 3.4 has not been tapped in all its generality. Coincidentally, our method is to do the natural generalization of Sz\"oll\H{o}si's idea to complex Hadamard matrices. The advantage of our method can be appreciated even for constructing families stemming from real Hadamard matrices. In these cases, we can apply our method to ER pairs of rows and columns simultaneously. Consequently, we are able to construct families stemming from real Hadamard matrices having more than $d/2+1$ independent parameters. For example, in dimension eight and twelve our method extends the families found by using Sz\"oll\H{o}si's method, as we will show further along in Section \ref{RHadamard}.

In order to construct a family of complex Hadamard matrices we can consider Di\c{t}\u{a}'s construction \cite{Dita}. This method can be applied to \emph{Di\c{t}\u{a} type} complex Hadamard matrices:\vspace{0.2cm}

\textbf{Di\c{t}\u{a} type}\hspace{0.2cm} \emph{A complex Hadamard matrix $H$ of order $d=d_1d_2$ is called Di\c{t}\u{a} type if there exists complex Hadamard matrices $M$ of order $d_1$ and  $N_1,\dots,N_{d_1}$ of order $d_2$ such that $H$ can be cast in the form}
\begin{equation}
    H=
\left( \begin{array}{ccc}
m_{11}N_1&\dots&m_{1d_1}N_{d_1}\\
\vdots&&\vdots\\
m_{d_11}N_1&\dots&m_{d_1d_1}N_{d_1}
\end{array} \right).
\end{equation}
The matrices $N_1,\dots,N_{d_1}$ are not necessarily different.

\textbf{Di\c{t}\u{a}'s construction}\hspace{0.2cm} \emph{Let $H$ be a Di\c{t}\u{a} type complex Hadamard matrix. Then, the following affine family stems from $H$}

\begin{equation}
    H(\vec{\xi})=
\left( \begin{array}{cccc}
m_{11}N_1&m_{12}D_2N_2&\dots&m_{1d_1}D_{d_1}N_{d_1}\\
\vdots&&&\vdots\\
m_{d_11}N_1&m_{d_12}D_2N_2&\dots&m_{d_1d_1}D_{d_1}N_{d_1}
\end{array} \right),
\end{equation}
where $D_2,\dots,D_{d_1}$ are diagonal unitary matrices and each of them contains $d_2-1$ free parameters. The total number of free parameters in this construction is $(d_1-1)(d_2-1)+m+n_1+\dots+n_{d_1}$. Here, $m$ and $n_1,\dots,n_{d_1}$ denote the number of free parameters of $M$ and $N_1,\dots,N_{d_1}$, respectively. In the case of  $N_1=N_2=\dots=N_{d_1}=N$ we have $H=M\otimes N$ and Di\c{t}\u{a} type matrices are reduced to Sylvester type \cite{Sylvester}. Let us show that our method intersects Di\c{t}\u{a}'s construction.
\begin{prop}\label{Dita_type}
Let $H$ be a dephased complex Hadamard matrix such that $\eta_{\bar{c}}=\eta_{\bar{r}}=d/2$. Then, $H$ is a Sylvester type.
\end{prop}
Proof: Suppose that there exists a complex Hadamard matrix such that $\eta_{\bar{c}}=\eta_{\bar{r}}=d/2$. Therefore, permuting rows and columns of $H$ we obtain the following equivalent matrix
\begin{equation}
    \tilde{H}=
\left( \begin{array}{cc}
A&B\\
A&-B
\end{array} \right),
\end{equation}
where $A$ and $B$ are $d/2\times d/2$ complex Hadamard matrices. Thus, our method is reduced to a particular case of Di\c{t}\u{a}'s construction
\begin{equation}\label{Dita_constr}
    H(\vec{\xi})=
\left( \begin{array}{cc}
A&D_1B\\
A&-D_1B
\end{array} \right)
\end{equation}
\qed

We emphasize that it is not easy to identify Di\c{t}\u{a} type matrices but it is straightforward to identify ER pairs even in higher dimensions.

\section{Fourier families in even dimensions}\label{Fourier}
In this section, we construct families of complex Hadamard matrices stemming from the Fourier matrices using our method. We analyze even $(d=2k)$ and doubly even $(d=4k)$ dimensions separately because different properties appear. We do not present new results in this section because the Fourier matrices have $\eta_{\bar{c}}=\eta_{\bar{r}}=d/2$ and then our method coincides with Di\c{t}\u{a}'s construction, as we have noticed in Proposition \ref{Dita_type}. We firstly analyze even dimensions.
\subsection{Fourier families in even dimensions}
In dimension four, the Fourier matrix is given by
\begin{equation}\label{F4}
    F_4=
\left( \begin{array}{lccc}
1&1&1&1\\
1&i&-1&-i\\
1&-1&1&-1\\
1&-i&-1&i
\end{array} \right).
\end{equation}
Here we have $\eta_c=\eta_r=2$ but only one independent parameter can be found after dephasing the family. Let $(C_2)_j$ and $(C_4)_j$ be the $j$th entries of the second and the fourth column of $F_4$, respectively. According to our mnemic technique defined after Eq.(\ref{C1C2}) we add a phase $e^{i\xi}$ in every entry of $C_1$ and $C_2$ if ${(C_2)_j}^*(C_4)_j=-1$. Consequently, we obtain the following family
\begin{equation}\label{FamilyF4}
    F_4^{(1)}(\xi)=F_4\circ\exp(iR_{F_4^{(1)}}(\xi)),
\end{equation}
where,
\begin{equation}
    R_{F_4^{(1)}}(\xi)=
\left( \begin{array}{lccc}
\bullet\hspace{0.2cm}&\bullet\hspace{0.2cm}&\bullet\hspace{0.2cm}&\bullet\\
\bullet&\xi&\bullet&\xi\\
\bullet&\bullet&\bullet&\bullet\\
\bullet&\xi&\bullet&\xi
\end{array} \right).
\end{equation}
Here, the symbol $\bullet$ means zero and $\xi\in[0,2\pi)$. This family is self-cognate and it agrees with the only maximal family existing in dimension four \cite{Bruzda}. In the same way, we construct a 2-parametric family stemming from the Fourier matrix in dimension six by considering the ER pairs of columns $\{C_2,C_5\}$ and $\{C_3,C_6\}$. That is,
\begin{equation}\label{FamilyF61}
    F_6^{(2)}(a,b)=F_6\circ\exp(iR^{(2)}_{F_6}(a,b)),
\end{equation}
and
\begin{equation}\label{FamilyF62}
    \left(F_6^{(2)}(a,b)\right)^t=F_6\circ\exp\left(i\left(R^{(2)}_{F_6}(a,b)\right)^t\right),
\end{equation}
where
\begin{equation}
    R_{F_6^{(2)}}(a,b)=
\left( \begin{array}{lccccc}
\bullet\hspace{0.2cm}&\bullet\hspace{0.2cm}&\bullet\hspace{0.2cm}&\bullet\hspace{0.2cm}&\bullet\hspace{0.2cm}&\bullet\\
\bullet&a&b&\bullet&a&b\\
\bullet&\bullet&\bullet&\bullet&\bullet&\bullet\\
\bullet&a&b&\bullet&a&b\\
\bullet&\bullet&\bullet&\bullet&\bullet&\bullet\\
\bullet&a&b&\bullet&a&b
\end{array} \right).
\end{equation}
This is the only maximal affine family stemming from the Fourier matrix $F_6$. As we have shown, the Fourier matrices contain $d/2$ ER pairs of columns in every even dimension. Therefore, we can construct the following affine families
\begin{equation}\label{mainsol_a}
    F_d^{(d/2-1)}(\vec{\xi}\,)=F_d\circ\exp(iR^{(d/2-1)}(\vec{\xi}\,)),
\end{equation}
and
\begin{equation}\label{mainsol_b}
    \left(F_d^{(d/2-1)}(\vec{\xi}\,)\right)^t=F_d\circ\exp\left(i\left(R^{(d/2-1)}(\vec{\xi}\,)\right)^t\right),
\end{equation}
where
\begin{equation}\label{mainsol_c}
    \left(R^{(d/2-1)}(\vec{\xi}\,)\right)_{i,j}=\left\{ \begin{array}{l}
\bullet\mbox{ when $i=0$ or even}\\
\bullet\mbox{ when $i$ is odd and }j=0\,\,\mathrm{mod}\,(d/2)\\
\xi_{[j-1]}\mbox{ otherwise.} \end{array} \right.
\end{equation}
The above families are not self-cognate. We have noticed from the BTZ catalog \cite{Bruzda} that in the cases $d=2,4,6,10,14$ our construction agrees with the maximal affine Hadamard family stemming from $F_d$. This motivates us to establish the following conjecture:
\begin{conje}
The maximal affine family of complex Hadamard matrices stemming from the Fourier matrix $F_d$ in dimensions $d=2p$ ($p$ prime) is given by Eqs.(\ref{mainsol_a}) to (\ref{mainsol_c}).
\end{conje}

\subsection{Fourier families in doubly even dimensions}\label{d4k}
In the case of dimensions of the form $d=4k,\,k>1$ we can simultaneously apply Theorem \ref{prop1} to rows and columns of the Fourier matrices. This procedure increases the number of parameters of the family beyond $d/2-1$. For example, we have obtained the following 5-parametric self-cognate family stemming from $F_8$
\begin{equation}\label{F8}
    F_8^{(5)}(a,b,c,d,e)=F_8\circ\exp(iR_{F_8}(a,b,c,d,e)),
\end{equation}
where
\begin{equation}\label{RF8}
    R_{F_8}(a,b,c,d,e)=
\left( \begin{array}{lccccccc}
\bullet\hspace{0.2cm}&\bullet\hspace{0.2cm}&\bullet\hspace{0.2cm}&\bullet\hspace{0.2cm}&\bullet\hspace{0.2cm}&\bullet\hspace{0.2cm}&\bullet\hspace{0.2cm}&\bullet\hspace{0.2cm}\\ \vspace{-0.1cm}
\bullet&a+d&e&a&\bullet&a+d&e&a\\ \vspace{-0.1cm}
\bullet&b&\bullet&b&\bullet&b&\bullet&b\\ \vspace{-0.1cm}
\bullet&c+d&e&c&\bullet&c+d&e&c\\ \vspace{-0.1cm}
\bullet&\bullet&\bullet&\bullet&\bullet&\bullet&\bullet&\bullet\\ \vspace{-0.1cm}
\bullet&a+d&e&a&\bullet&a+d&e&a\\ \vspace{-0.1cm}
\bullet&b&\bullet&b&\bullet&b&\bullet&b\\ \vspace{-0.1cm}
\bullet&c+d&e&c&\bullet&c+d&e&c
\end{array} \right).
\end{equation}
The ER pairs here considered are $\{C_2,C_6\},\{C_3,C_7\},\{C_4,C_8\}$ and $\{R_2,R_6\},\{R_3,R_7\},\{R_4,R_8\}$ ($C=R=\{2,6;3,7;4,8\}$ to abbreviate). The ER pair $\{C_4,C_8\}$ produces a linearly dependent parameter and it has not been considered in Eq.(\ref{RF8}). The 5-parametric family given in Eq.(\ref{F8}) coincides with the maximal affine family stemming from $F_8$ \cite{Tadej2}. In the case of $d=12$, the following 9-parametric family can be found
\begin{equation}\label{FamilyF12}
    F_{12}^{(9)}(a,b,c,d,e,f,g,h,i)=F_{12}\circ\exp(iR_{F_{12}}(a,b,c,d,e,f,g,h,i)),
\end{equation}
where
\begin{equation}
    R_{F_{12}}(a,b,c,d,e,f,g,h,i)=
\left( \begin{array}{lccccccccccc}
\bullet\hspace{0.2cm}&\bullet\hspace{0.2cm}&\bullet\hspace{0.2cm}&\bullet\hspace{0.2cm}&\bullet\hspace{0.2cm}&\bullet\hspace{0.2cm}&\bullet\hspace{0.2cm}&\bullet\hspace{0.2cm}&\bullet\hspace{0.2cm}&\bullet\hspace{0.2cm}&\bullet\hspace{0.2cm}&\bullet\\
\bullet&a+f&g&a+h&i&a&\bullet&a+f&g&a+h&i&a\\
\bullet&b&\bullet&b&\bullet&b&\bullet&b&\bullet&b&\bullet&b\\
\bullet&c+f&g&c+h&i&c&\bullet&c+f&g&c+h&i&c\\
\bullet&d&\bullet&d&\bullet&d&\bullet&d&\bullet&d&\bullet&d\\
\bullet&e+f&g&e+h&i&e&\bullet&e+f&g&e+h&i&e\\
\bullet\hspace{0.2cm}&\bullet\hspace{0.2cm}&\bullet\hspace{0.2cm}&\bullet\hspace{0.2cm}&\bullet\hspace{0.2cm}&\bullet\hspace{0.2cm}&\bullet\hspace{0.2cm}&\bullet\hspace{0.2cm}&\bullet\hspace{0.2cm}&\bullet\hspace{0.2cm}&\bullet\hspace{0.2cm}&\bullet\\
\bullet&a+f&g&a+h&i&a&\bullet&a+f&g&a+h&i&a\\
\bullet&b&\bullet&b&\bullet&b&\bullet&b&\bullet&b&\bullet&b\\
\bullet&c+f&g&c+h&i&c&\bullet&c+f&g&c+h&i&c\\
\bullet&d&\bullet&d&\bullet&d&\bullet&d&\bullet&d&\bullet&d\\
\bullet&e+f&g&e+h&i&e&\bullet&e+f&g&e+h&i&e
\end{array} \right).
\end{equation}
Here, we consider the ER pairs $C=R=\{2,8;3,9;4,10;5,11;6,12\}$, whereas $\{C_6,C_{12}\}$ produces a linearly dependent parameter. This result coincides with the family $F_{12A}^{(9)}$ \cite{Tadej2} which is self-cognate. A non-affine family stemming from $F_{12}$ has been recently found by Barros and Bengtsson \cite{Barros}. This family contains the affine families found by Tadej and \.{Z}yczkowski \cite{Tadej2}. Finally, we obtain a 13-parametric family stemming from $F_{16}$. That is,
\begin{equation}
    F_{16}^{(13)}(a,b,c,d,e,f,g,h,i,j,k,l,m)=F_{16}\circ\exp(iR_{F_{16}}(a,b,c,d,e,f,g,h,i,j,k,l,m)),
\end{equation}
where $R_{F_{16}}(a,b,c,d,e,f,g,h,i,j,k,l,m)$ is given by
\begin{equation}
\left( \begin{array}{lccccccccccccccc}
\bullet\hspace{0.2cm}&\bullet\hspace{0.2cm}&\bullet\hspace{0.2cm}&\bullet\hspace{0.2cm}&\bullet\hspace{0.2cm}&\bullet\hspace{0.2cm}&\bullet\hspace{0.2cm}&\bullet\hspace{0.2cm}&\bullet\hspace{0.2cm}&\bullet\hspace{0.2cm}&\bullet\hspace{0.2cm}&\bullet&\bullet\hspace{0.2cm}&\bullet\hspace{0.2cm}&\bullet\hspace{0.2cm}&\bullet\\
\bullet&a+h&i&a+j&k&a+l&m&a&\bullet&a+h&i&a+j&k&a+l&m&a\\
\bullet&b&\bullet&b&\bullet&b&\bullet&b&\bullet&b&\bullet&b&\bullet&b&\bullet&b\\
\bullet&c+h&i&c+j&k&c+l&m&c&\bullet&c+h&i&c+j&k&c+l&m&c\\
\bullet&d&\bullet&d&\bullet&d&\bullet&d&\bullet&d&\bullet&d&\bullet&d&\bullet&d\\
\bullet&e+h&i&e+j&k&e+l&m&e&\bullet&e+h&i&e+j&k&e+l&m&e\\
\bullet&f&\bullet&f&\bullet&f&\bullet&f&\bullet&f&\bullet&f&\bullet&f&\bullet&f\\
\bullet&g+h&i&g+j&k&g+l&m&g&\bullet&g+h&i&g+j&k&g+l&m&g\\
\bullet\hspace{0.2cm}&\bullet\hspace{0.2cm}&\bullet\hspace{0.2cm}&\bullet\hspace{0.2cm}&\bullet\hspace{0.2cm}&\bullet\hspace{0.2cm}&\bullet\hspace{0.2cm}&\bullet\hspace{0.2cm}&\bullet\hspace{0.2cm}&\bullet\hspace{0.2cm}&\bullet\hspace{0.2cm}&\bullet&\bullet\hspace{0.2cm}&\bullet\hspace{0.2cm}&\bullet\hspace{0.2cm}&\bullet\\
\bullet&a+h&i&a+j&k&a+l&m&a&\bullet&a+h&i&a+j&k&a+l&m&a\\
\bullet&b&\bullet&b&\bullet&b&\bullet&b&\bullet&b&\bullet&b&\bullet&b&\bullet&b\\
\bullet&c+h&i&c+j&k&c+l&m&c&\bullet&c+h&i&c+j&k&c+l&m&c\\
\bullet&d&\bullet&d&\bullet&d&\bullet&d&\bullet&d&\bullet&d&\bullet&d&\bullet&d\\
\bullet&e+h&i&e+j&k&e+l&m&e&\bullet&e+h&i&e+j&k&e+l&m&e\\
\bullet&f&\bullet&f&\bullet&f&\bullet&f&\bullet&f&\bullet&f&\bullet&f&\bullet&f\\
\bullet&g+h&i&g+j&k&g+l&m&g&\bullet&g+h&i&g+j&k&g+l&m&g
\end{array} \right).
\end{equation}
This family is self-cognate. The ER pairs here considered are $C=R=\{2,10;3,11;4,12;5,13;6,14;7,15\}$, whereas $\{C_8,C_{16}\}$ produces a linearly dependent parameter. We remark that the maximal affine family stemming from $F_{16}$ has 17 parameters \cite{Dita,Tadej2}. In general, for $d=4k,\,k>1$ we can construct the following families using our method:
\begin{equation}\label{mainsol2_a}
    F_d(\vec{\alpha},\vec{\beta})=F_d\circ\exp(i(R_1(\vec{\alpha})+R_2(\vec{\beta}))),
\end{equation}
and
\begin{equation}\label{mainsol2_b}
    \left(F_d(\vec{\alpha},\vec{\beta})\right)^T=F_d\circ\exp\left(i\left(R_1(\vec{\alpha})+R_2(\vec{\beta})\right)^T\right),
\end{equation}
where
\begin{equation}
    \left(R_1(\vec{\alpha})\right)_{i,j}=\left\{ \begin{array}{l}
\bullet\mbox{ when $i=0$ or even}\\
\bullet\mbox{ when $i$ is odd and }j=0\,\,\mathrm{mod}\,(d/2)\\
\alpha_{[j-1]}\mbox{ otherwise,} \end{array} \right.
\end{equation}
and
\begin{equation}
    R_2(\vec{\beta})=R^T_1(\vec{\beta}).
\end{equation}
Note that $\vec{\alpha},\vec{\beta}$ can be dependent and $d-2$ is an upper bound for the dimension of the family that we can obtain from $F_d$.

We can define a lower bound for the maximal affine family stemming from a complex Hadamard matrix in even dimensions.
\begin{prop}
Let $H$ be a complex Hadamard matrix and $d_{max}(H)$ the dimension of the maximal affine family stemming from $H$. Then, the following lower bound can be established
\begin{equation}
    \eta_{max}(H)\leq d_{max}(H).
\end{equation}
\end{prop}
where $\eta_{max}$ is the maximal number of linearly independent parameters of a family constructed by using our method. Let us analyze this bound in the case of the Fourier matrices in even dimensions.
\begin{center}
\begin{tabular}{|c|c|c||c|c|c|}
  \hline
\hspace{0.2cm}$d$\hspace{0.2cm} & $\eta_{max}(F_d)$ & $d_{max}(F_d)$ & \hspace{0.2cm}$d$\hspace{0.2cm} & $\eta_{max}(F_d)$ & $d_{max}(F_d)$ \\ \hline
2&0&0&10&4&4\\
4&1&1&12&9&9\\
6&2&2&14&6&6\\
8&5&5&16&13&17\\
  \hline
\end{tabular}
\end{center}
As we can see in this table, up to $d=14$ every known affine family stemming from the Fourier matrix can be constructed by using our method.

\section{Real Hadamard matrices}\label{RHadamard}
Fourier matrices have $d/2$ aligned ER pairs of rows and columns for every even dimension. However, for real Hadamard matrices the ER pairs are not necessarily aligned. Also, we have many non-equivalent ways to define the ER pairs. As we will show next, for $d\geq8$ the ER pairs should be intelligently chosen in order to maximize the dimension of a family. In $d=2$ we have an isolated real Hadamard matrix. For $d=4$ every real Hadamard matrix belongs to the one parametric family $F_4^{(1)}(\xi)$ presented in the previous section. Thus, our first interesting case is $d=8$. Here, every real Hadamard matrix is equivalent to
\begin{equation}
    H_8=
\left( \begin{array}{rrrrrrrr}
1&1&1&1&1&1&1&1\\
1&1&-1&1&-1&-1&1&-1\\
1&1&1&-1&-1&-1&-1&1\\
1&-1&1&1&-1&1&-1&-1\\
1&-1&1&-1&1&-1&1&-1\\
1&-1&-1&1&1&-1&-1&1\\
1&1&-1&-1&1&1&-1&-1\\
1&-1&-1&-1&-1&1&1&1
\end{array} \right).
\end{equation}
In order to maximize a family obtained from $H_8$, we should choose the ER pairs of rows such that the number of ER pairs of columns is maximal. Let us explicitly construct the family. Applying our method to the rows of $H_8$ and considering $\eta_{\bar{r}}=4$ we obtain, as a first step, the 4-parametric family
\begin{equation}\label{H8}
    H_8(a,b,c,d)=H_8\circ\exp(iR_{H_8}(a,b,c,d)),
\end{equation}
where
\begin{equation}\label{RH8}
R_{H_8}(a,b,c,d)=
\left( \begin{array}{llllllll}
\bullet\hspace{0.2cm}&a\hspace{0.2cm}&\bullet\hspace{0.2cm}&a\hspace{0.2cm}&\bullet\hspace{0.2cm}&a\hspace{0.2cm}&\bullet\hspace{0.2cm}&a\\
\bullet&b&\bullet&b&\bullet&b&\bullet&b\\
\bullet&c&\bullet&c&\bullet&c&\bullet&c\\
\bullet&c&\bullet&c&\bullet&c&\bullet&c\\
\bullet&a&\bullet&a&\bullet&a&\bullet&a\\
\bullet&d&\bullet&d&\bullet&d&\bullet&d\\
\bullet&d&\bullet&d&\bullet&d&\bullet&d\\
\bullet&b&\bullet&b&\bullet&b&\bullet&b
\end{array} \right).
\end{equation}
Therefore, we have $9$ inequivalent ways to choose the ER pairs of columns. They are:
\begin{eqnarray}
   C_A&=&\{1,3;5,7;2,4;6,8\},\\
   C_B&=&\{1,3;5,7;2,6;4,8\},\\
   C_C&=&\{1,3;5,7;2,8;4,6\},\\
   C_D&=&\{1,5;3,7;2,4;6,8\},\\
   C_E&=&\{1,5;3,7;2,6;4,8\},\\
   C_F&=&\{1,5;3,7;2,8;4,6\},\\
   C_G&=&\{1,7;3,5;2,4;6,8\},\\
   C_H&=&\{1,7;3,5;2,6;4,8\},\\
   C_I&=&\{1,7;3,5;2,8;4,6\}.\label{choices}
\end{eqnarray}
Let us analyze the case of $C_A$. That is,
\begin{equation}
    H^{(8)}_{8A}(a,b,c,d,e)=H_8\circ\exp(iR_{H_{8A}}(a,b,c,d,e)),
\end{equation}
where
\begin{equation}
R_{H_{8A}}(a,b,c,d,e)=
\left( \begin{array}{lccccccc}
\bullet\hspace{0.2cm}&\bullet\hspace{0.2cm}&\bullet\hspace{0.2cm}&\bullet\hspace{0.2cm}&\bullet\hspace{0.2cm}&\bullet\hspace{0.2cm}&\bullet\hspace{0.2cm}&\bullet\hspace{0.2cm}\\
\bullet&a+b+d&\bullet&a+b+d&d&a+b+d&d&a+b+d\\
\bullet&a+c+e&\bullet&a+c+e&\bullet&a+c&\bullet&a+c\\
\bullet&a+c+e&\bullet&a+c+e&\bullet&a+c&\bullet&a+c\\
\bullet&\bullet&\bullet&\bullet&\bullet&\bullet&\bullet&\bullet\\
\bullet&a+d+e&\bullet&a+d+e&d&a+d&d&a+d\\
\bullet&a+d+e&\bullet&a+d+e&d&a+d&d&a+d\\
\bullet&a+b+d&\bullet&a+b+d&d&a+b+d&d&a+b+d
\end{array} \right).
\end{equation}
We can considerer $7!!=7\times5\times3\times1=105$ different choices for the ER pairs of rows. The above 9 cases consider only one of these choices. In principle, we do not know how many of these families are inequivalent. Using Di\c{t}\u{a}'s construction there were found 9 inequivalent 5-parametric families stemming from $H_8$ \cite{Dita2}, which are a particular subset of our solutions here presented. This is very easy to show from Proposition \ref{Dita_type}.

In Section \ref{Szol_Dita}, we already mentioned that our method generalizes Sz\"oll\H{o}si's method even for real Hadamard matrices. Let us present an example where this is clearly showed. From the real Hadamard matrix
\begin{equation}\label{H12}
H_{12}=
\left( \begin{array}{rrrrrrrrrrrr}
1&1&1&1&1&1&1&1&1&1&1&1\\ \vspace{0.1cm}
1&1&1&1&1&1&-1&-1&-1&-1&-1&-1\\ \vspace{0.1cm}
1&-1&1&1&-1&-1&1&1&-1&1&-1&-1\\ \vspace{0.1cm}
1&-1&1&-1&1&-1&-1&1&1&-1&1&-1\\ \vspace{0.1cm}
1&-1&-1&1&-1&1&-1&1&1&-1&-1&1\\ \vspace{0.1cm}
1&1&1&-1&-1&-1&1&-1&1&-1&-1&1\\ \vspace{0.1cm}
1&1&-1&-1&1&-1&-1&1&-1&1&-1&1\\ \vspace{0.1cm}
1&1&-1&1&-1&-1&-1&-1&1&1&1&-1\\ \vspace{0.1cm}
1&-1&1&-1&-1&1&-1&-1&-1&1&1&1\\ \vspace{0.1cm}
1&1&-1&-1&-1&1&1&1&-1&-1&1&-1\\ \vspace{0.1cm}
1&-1&-1&-1&1&1&1&-1&1&1&-1&-1\\ \vspace{0.1cm}
1&-1&-1&1&1&-1&1&-1&-1&-1&1&1
\end{array} \right),
\end{equation}
we find an 8-parametric family from considering the ER pairs given by $C=\{1,6;2,3\}$ and $R=\{1,5;2,11;3,7;4,8;6,12;9,10\}$. That is,
\begin{equation}\label{FamilyF12R}
    H^{(8)}_{12}(a,b,c,d,e,f,g,h)=H_{12}\circ\exp(iR_{H_{12}}(a,b,c,d,e,f,g,h)),
\end{equation}
where $R_{H_{12}}(a,b,c,d,e,f,g,h)$ is given by
\begin{equation}\label{FamilyF12R2}
\left( \begin{array}{lccccccccccc}
\bullet\hspace{0.2cm}&\bullet\hspace{0.2cm}&\bullet\hspace{0.2cm}&\bullet\hspace{0.2cm}&\bullet\hspace{0.2cm}&\bullet\hspace{0.2cm}&\bullet\hspace{0.2cm}&\bullet\hspace{0.2cm}&\bullet\hspace{0.2cm}&\bullet\hspace{0.2cm}&\bullet\hspace{0.2cm}&\bullet\hspace{0.2cm}\\
\bullet&a+b&a+b&b&a&\bullet&a+b&\bullet&b&a+b&a&\bullet\\
\bullet&a+c\atop+g+h&a+c\atop+g+h&c+h&a+c\atop+h&\bullet&a+c\atop+h&h&h&a+h&a+h&c+h\\
\bullet&a+d\atop+g+h&a+d\atop+g+h&d+h&a+d\atop+h&\bullet&a+h&d+h&h&a+d\atop+h&a+h&h\\
\bullet\hspace{0.2cm}&\bullet\hspace{0.2cm}&\bullet\hspace{0.2cm}&\bullet\hspace{0.2cm}&\bullet\hspace{0.2cm}&\bullet\hspace{0.2cm}&\bullet\hspace{0.2cm}&\bullet\hspace{0.2cm}&\bullet\hspace{0.2cm}&\bullet\hspace{0.2cm}&\bullet\hspace{0.2cm}&\bullet\hspace{0.2cm}\\
\bullet&a+e\atop+h&a+e\atop+h&e+h&a+e\atop+h&\bullet&a+h&h&e+h&a+h&a+e\atop+h&h\\
\bullet&a+c\atop+g+h&a+c\atop+g+h&c+h&a+c\atop+h&\bullet&a+c\atop+h&h&h&a+h&a+h&c+h\\
\bullet&a+d\atop+g+h&a+d\atop+g+h&d+h&a+d\atop+h&\bullet&a+h&d+h&h&a+d\atop+h&a+h&h\\
\bullet&a+f\atop+g&a+f\atop+g&\bullet&a&\bullet&a+f&f&\bullet&a+f&a&f\\
\bullet&a+f\atop+g&a+f\atop+g&\bullet&a&\bullet&a+f&f&\bullet&a+f&a&f\\
\bullet&a+b&a+b&b&a&\bullet&a+b&\bullet&b&a+b&a&\bullet\\
\bullet&a+e\atop+h&a+e\atop+h&e+h&a+e\atop+h&\bullet&a+h&h&e+h&a+h&a+e+h&h
\end{array} \right).
\end{equation}
We have verified that all the parameters are linearly independent. Considering different combinations of ER pairs of rows and columns we can generate $12!!=46,080$ families but we do not know how many of them are inequivalent. Our family $H^{(8)}_{12}$ is a new result that extends the family $H^{(7)}_{12}$ found by using Sz\"oll\H{o}si's method \cite{Szollosi}.

We have noted that our method can be applied to other complex Hadamard matrices apart from the real and the Fourier matrices. For example, from $H^{(\omega)}_{10}$ \cite{Dita4} we found the following 5-parametric family
\begin{equation}\label{Hw}
    H^{(\omega)}_{10}(a,b,c,d,e)=H^{(\omega)}_{10}\circ\exp(iR_{H^{(\omega)}_{10}}(a,b,c,d,e)),
\end{equation}
where
\begin{equation}
H^{(\omega)}_{10}=
\left( \begin{array}{rrrrrrrrrr}
1&1&1&1&1&1&1&1&1&1\\  \vspace{-0.1cm}
1&1&1&\omega&\omega^2&-1&-1&1&\omega^2&\omega\\  \vspace{-0.1cm}
1&1&1&\omega^2&\omega&-1&1&-1&\omega&\omega^2\\ \vspace{-0.1cm}
1&\omega&\omega^2&1&1&-1&\omega^2&\omega&-1&1\\ \vspace{-0.1cm}
1&\omega^2&\omega&1&1&-1&\omega&\omega^2&1&-1\\ \vspace{-0.1cm}
1&\omega^2&\omega&1&-1&1&-\omega&-\omega^2&-1&-1\\ \vspace{-0.1cm}
1&-1&1&\omega&\omega^2&1&-1&-1&-\omega^2&-\omega\\ \vspace{-0.1cm}
1&1&-1&\omega^2&\omega&1&-1&-1&-\omega&-\omega^2\\ \vspace{-0.1cm}
1&\omega&\omega^2&-1&1&1&-\omega^2&-\omega&-1&-1\\ \vspace{-0.1cm}
1&-1&-1&-1&-1&-1&1&1&1&1
\end{array} \right),
\end{equation}
$\omega^2+\omega+1=0$ and
\begin{equation}\label{Romega10}
R_{H^{(\omega)}_{10}}(a,b,c,d,e)=
\left( \begin{array}{lccccccccc}
\bullet\hspace{0.2cm}&\bullet\hspace{0.2cm}&\bullet\hspace{0.2cm}&\bullet\hspace{0.2cm}&\bullet\hspace{0.2cm}&\bullet\hspace{0.2cm}&\bullet\hspace{0.2cm}&\bullet\hspace{0.2cm}&\bullet\hspace{0.2cm}&\bullet\hspace{0.2cm}\\
\bullet&a+b&a&a&a&a+b&\bullet&b&b&b\\
\bullet&a&a+c&a&a&a+c&c&\bullet&c&c\\
\bullet&a&a&a+d&a&a+d&d&d&\bullet&d\\
\bullet&a&a&a&a+e&a+e&e&e&e&\bullet\\
\bullet&a&a&a&a+e&a+e&e&e&e&\bullet\\
\bullet&a+b&a&a&a&a+b&\bullet&b&b&b\\
\bullet&a&a+c&a&a&a+c&c&\bullet&c&c\\
\bullet&a&a&a+d&a&a+d&d&d&\bullet&d\\
\bullet\hspace{0.2cm}&\bullet\hspace{0.2cm}&\bullet\hspace{0.2cm}&\bullet\hspace{0.2cm}&\bullet\hspace{0.2cm}&\bullet\hspace{0.2cm}&\bullet\hspace{0.2cm}&\bullet\hspace{0.2cm}&\bullet\hspace{0.2cm}&\bullet\hspace{0.2cm}
\end{array} \right).
\end{equation}
In this case, we have $\eta_c=5$ and $\eta_r=5$ but they cannot be simultaneously considered in order to obtain more free parameters. It was proven that a 7-parametric family stems from $H^{(\omega)}_{10}$ \cite{Dita4}, and Eq.(\ref{Hw}) represents a subset of this family. We have also extended the 7 parametric family $D^{(7)}_{12\Sigma}$ \cite{Dita3} found by using Di\c{t}\u{a}'s construction. This extension is obtained by considering the ER pair given by the first and the last row in every subfamily of $D^{(7)}_{12\Sigma}$. This is straightforwardly obtained by inspecting of the family \cite{Dita3}. We have proven that the new parameter is linearly independent to the rest of the parameters after dephasing the family. We have omitted details here to abbreviate but the twenty subfamilies $D^{(7)}_{12A}$ to $D^{(7)}_{12R}$ are generalized in the same straightforward way. As a last result, we show that the single matrix $D_{12}$ presented by Di\c{t}\u{a} \cite{Dita4}
 \begin{equation}
D_{12}=
\left( \begin{array}{rrrrrrrrrrrr}
1&1&1&1&1&1&1&1&1&1&1&1\\
1&i&i&i&-i&-i&-i&-1&1&1&-1&-1\\
1&i&i&-i&i&-i&-i&1&-1&-1&1&-1\\
1&i&-i&i&-i&i&-i&1&-1&-1&-1&1\\
1&-i&i&-i&i&i&-i&-1&1&-1&-1&1\\
1&-i&-i&i&i&i&-i&-1&-1&1&1&-1\\
1&-i&-i&i&i&-i&i&1&1&-1&-1&-1\\
1&-i&i&i&-i&-i&i&-1&-1&-1&1&1\\
1&i&-i&-i&i&-i&i&-1&-1&1&-1&1\\
1&i&-i&-i&-i&i&i&-1&1&-1&1&-1\\
1&-i&i&-i&-i&i&i&1&-1&1&-1&-1\\
1&-1&-1&-1&-1&-1&-1&1&1&1&1&1
\end{array} \right),
\end{equation}
belongs to the intersection of our extension of $D^{(7)}_{12\Sigma}$, namely $D^{(8)}_{12\Sigma}$. That is,
\begin{equation}\label{D12}
    D_{12}\in\bigcap_{\Gamma=A}^R D^{(8)}_{12\Gamma}.
\end{equation}
Indeed, every subfamily of $D^{(7)}_{12\Sigma}$ stems from a real Hadamard matrix equivalent to $H_{12}$. All these equivalent real matrices have the same first and last row as $H_{12}$. In order to obtain $D_{12}$ we start by multiplying from the second to the seventh column of $H_{12}$ by $i$ times. Thus, we introduce a parameter $\xi$ by applying our method to the ER pair $\{R_1,R_{12}\}$. In the case of $\xi=-\pi/2$ we obtain $D_{12}$. Analogously for every subfamily $D^{(8)}_{12\Gamma},\,\Gamma=A,\dots,R$.

\section{Summary and conclusion}
We presented a new method to construct families of complex Hadamard matrices in every even dimension $d>2$ by introducing the concept of \emph{ER pairs}. Let us summarize our results:

\noindent\emph{(i)} Using our method we have reproduced some previously known results:

\noindent\emph{We found maximal affine families stemming from the Fourier matrix in $d=4$ (Eq.(\ref{FamilyF4})), $6$ (Eqs.(\ref{FamilyF61}) and (\ref{FamilyF62})), $8$ (Eq.(\ref{F8})) and $12$ (Eq.(\ref{FamilyF12})). Also, we found families stemming from the Fourier matrix in every even (Eqs.(\ref{mainsol_a}) and (\ref{mainsol_b})) and double even (Eqs.(\ref{mainsol2_a}) and (\ref{mainsol2_b})) dimension.}\\ \vspace{-0.5cm}

\noindent \emph{(ii)} Although our method is defined in even dimensions, we also found a restriction on the distribution of a parameter in families existing in every odd dimension:

\noindent\emph{Families of complex Hadamard matrices defined in every odd dimension cannot contain a parameter appearing in only two columns or rows (See Corollary \ref{oddcorol}).}\\ \vspace{-0.5cm}

\noindent \emph{(iii)} We have generalized Sz\"oll\H{o}si's method for constructing affine families:

\noindent\emph{Our method increases the number of free independent parameters that can be obtained by using Sz\"oll\H{o}si's method in every even dimension $d\geq12$. (See the beginning of Section \ref{Szol_Dita}).}\\ \vspace{-0.5cm}

\noindent \emph{(iv)} Our method has an intersection with Di\c{t}\u{a}'s construction:

\noindent \emph{If a $d\times d$ complex Hadamard matrix has $d/2$ aligned ER pairs of columns or rows then it is Di\c{t}\u{a} type. (See Proposition \ref{Dita_type}).}\\ \vspace{-0.5cm}

\noindent \emph{(v)} We have constructed several families stemming from $H_8$ in dimension eight:

\noindent \emph{We found $9\times105=945$ different ways to construct a 5-dimensional family stemming from the real Hadamard matrix $H_{8}$} (See Eq.(\ref{H8}) to Eq.(\ref{choices})).\\ \vspace{-0.5cm}

\noindent \emph{(vi)} We have extended two families in dimension twelve:

\noindent \emph{We found 46,080 different ways to generalize the family $H_{12}^{(7)}$ obtained from Sz\"oll\H{o}si's method} (See Eqs. (\ref{FamilyF12R})-(\ref{FamilyF12R2}) and the paragraph afterwards).\\ \vspace{-0.5cm}

\noindent \emph{The family $D_{12\Sigma}^{(7)}$ obtained from Di\c{t}\u{a}'s construction was extended to $D_{12\Sigma}^{(8)}$ (See paragraph after Eq.(\ref{Romega10})). Also, we have proven that the single matrix $D_{12}$ presented by Di\c{t}\u{a} belongs to every subfamily of $D_{12\Sigma}^{(8)}$} (See Eq.(\ref{D12})).\\ \vspace{-0.5cm}

\noindent \emph{(vii)} We have established a connection between the mutually unbiased (MU) bases problem in dimension six and the ER pairs:

\noindent\emph{Let $\{\mathbb{I},H_1,H_2,H_3\}$ be a set of four MU bases existing in dimension six. Then, $H_1,H_2$ and $H_3$ do \emph{not} have ER pairs} (See Corollary \ref{CorolMUB}).\\ \vspace{-0.5cm}

And finally,

\noindent \emph{(viii)} We generated inequivalent affine families stemming from inequivalent real Hadamard matrices. For example:

\noindent \emph{In dimensions 16, 20, 24, 28 and 32 we can construct 5, 3, 60, 487 and more than 13 millon inequivalent families, respectively} (See Proposition \ref{propHad} and Corollary \ref{Corol13millon}).\\

Our method to construct families considers parameters appearing in pairs of columns or rows of complex Hadamard matrices. This assumption allowed us to construct many families of complex Hadamard matrices in a very easy way. However, our method is not general because the parameters can appear in more than two columns. In fact, several families existing in even dimensions and all families existing in every odd dimension cannot be constructed from our method. This naturally suggests to us to try to generalize the concept of ER pairs. Nevertheless, a general extension to three or four columns seems not easy. We have noted that parameters appearing in three columns can only be shown in the following families of the BTZ catalog: $F^{(2)}_6$, $F^{(4)}_9$, $S^{(5)}_{12}$. This evidence strongly suggests that parameters appearing in exactly three columns are only possible in dimensions of the form $d=3k$. Very interestingly, parameters appearing in four columns are not restricted to doubly even dimensions, as we can see in Petrescu's family $P^{(1)}_7$. Therefore, a generalization to four columns could lead us to a construction of affine families in every dimension $d\geq6$.

We have solved the problem of finding the maximal affine family stemming from a complex Hadamard matrix when the parameters appear in exactly two columns or rows. We hope this method and its generalization to be a useful tool to try to understand the general structure of affine families of complex Hadamard matrices existing in every dimension.

  \section{Acknowledgments}
I specially thank to W. Tadej, I. Bengtsson, K. \.{Z}yczkowski, S. Weigert, P. Di\c{t}\u{a}, A. Delgado, F. Sz\"oll\H{o}si, B. Karlsson and M. Matolcsi for their invaluable comments. Also, I would like to thank to the referee for his many useful comments in order to improve this article. This work is supported by Grants FONDECyT N$^{\text{\underline{o}}}$ 3120066 and MSI P010-30F.

\end{document}